\documentclass[10pt]{article}

\usepackage[margin=0.72in]{geometry}
\usepackage{amsmath,amssymb}
\usepackage{graphicx}
\usepackage{subcaption}
\usepackage{booktabs,tabularx}
\usepackage{microtype}
\usepackage{newtxtext,newtxmath}
\usepackage[backend=biber,style=authoryear,sorting=nyt,maxbibnames=6,minbibnames=3,maxcitenames=2,mincitenames=1,uniquelist=false]{biblatex}
\let\cite\parencite
\AtEveryBibitem{\clearfield{title}}
\usepackage[hidelinks]{hyperref}
\usepackage{placeins}

\graphicspath{{figures/}}

\newcommand{\sqmUnit}{\ensuremath{\mathrm{mag\,arcsec^{-2}}}}
\newcommand{\figref}[1]{\hyperref[#1]{Figure~\ref*{#1}}}
\newcommand{\figshortref}[1]{\hyperref[#1]{Fig.~\ref*{#1}}}
\newcommand{\tabref}[1]{\hyperref[#1]{Table~\ref*{#1}}}
\newcommand{\eqnref}[1]{\hyperref[#1]{Equation~(\ref*{#1})}}

\title{Low-Cost Streetlight Shields for Dark-Sky Protection at a Stargazing Camp}
\author{\parbox{\dimexpr\textwidth-2\tabcolsep\relax}{\raggedright\normalsize\textbf{Xiaoyong Liu}\textsuperscript{1}, \textbf{Ruobin Ding}\textsuperscript{2}, \textbf{Sixuan Zhang}\textsuperscript{3}\\[0.35\baselineskip]\textit{\textsuperscript{1} The Experimental High School Attached to Beijing Normal University, Beijing, China; leoliuxiaoyong06@outlook.com\\\textsuperscript{2} School of Physics and Astronomy, Beijing Normal University, Beijing 100875, China; dingruobin@mail.bnu.edu.cn\\\textsuperscript{3} Hiroshima Astrophysical Science Center, Hiroshima University, 1-3-1 Kagamiyama, Higashi-Hiroshima, Hiroshima 739-8526 Japan}}}
\date{}

\begin{document}
\maketitle

{\small
\noindent\textbf{Abstract}\par
\vspace{0.35\baselineskip}
\noindent Road lighting at stargazing camps can cause glare and stray light that interfere with observation. We installed directional shields on existing solar-powered streetlights at Bayan Gol Starlight Camp in Inner Mongolia, China, and compared SQM-L measurements before and after installation at the camp center (P1) and a site approximately 500~m away (P2). Measurements were taken at elevations of $0^\circ$, $30^\circ$, $60^\circ$, and $90^\circ$. At $0^\circ$, the mean SQM reading across four directions at P1 increased by 1.624~\sqmUnit{}, and the reading toward the camp at P2 increased by 2.312~\sqmUnit{}. These changes correspond to reductions of approximately 77.6\% and 88.1\% in the equivalent SQM signal within the measured fields of view. The largest changes occurred at $0^\circ$; changes at $30^\circ$ and above were smaller. Photographs taken with identical exposure settings showed less visible glare around the shielded lamps. The measurements indicate that the changes were concentrated in low-elevation views. Directional shields provide a relatively low-cost way to reduce unwanted light while retaining the existing road-lighting system.
\par}

\vspace{1.0\baselineskip}

\section{Introduction}

Artificial light scattered by the atmosphere creates skyglow and makes faint stars harder to see. A global atlas estimated that more than 80\% of people live under skies affected by artificial light~\cite{falchi2016}. Satellite observations showed increases in both the area and radiance of Earth's artificially lit surface from 2012 to 2016~\cite{kyba2017}. Based on public star-visibility reports, \textcite{kyba2023} estimated an average annual increase of about 9.6\% in sky brightness at participants' locations from 2011 to 2022. Artificial light at night can also affect animals' nighttime activity~\cite{longcore2004}.

At a stargazing camp, skyglow and direct glare are distinct problems. Skyglow raises the background brightness of the sky, while a visible lamp can cause glare and interfere with dark adaptation. A dark zenith reading does not rule out glare from a lamp near an entrance or road. Site assessments should include low-elevation views as well as the zenith~\cite{hanel2018}.

Roads and entrances at the camp still need lighting, so switching off all lamps may not meet site needs. Outdoor-lighting guidance recommends directing light only where it is needed and controlling its level, duration, and color~\cite{darksky2020}.

Possible measures include dimming lamps, shortening operating hours, using warmer light, or replacing fixtures with fully shielded models. We tested a smaller retrofit: shields fitted to existing streetlights to limit sideward and upward light. They fit the original lamp housing, left the underside open, and required no electrical changes. Compared with replacing complete fixtures or modifying the electrical system, this retrofit required fewer new materials and less installation work because the existing lamps, solar panels, poles, and control systems were retained. It therefore provides a relatively low-cost option for improving the nighttime light environment at small stargazing sites. This approach is consistent with guidance to limit obtrusive and spill light~\cite{gaston2012,cie2017,darksky2020}.

Light levels can vary within a site with lamp orientation, visibility of the light-emitting surface, and nearby buildings or terrain. \textcite{katz2016} found that SQM readings differ when the meter is aimed at the sky, ground, or horizon. A countywide LED retrofit study reported different trends in ground-based and satellite observations~\cite{hung2021}. We therefore measured several directions and elevations; photographs document visible glare but do not replace instrument readings.

\begin{figure}[!htbp]
  \centering
  \includegraphics[width=0.78\linewidth]{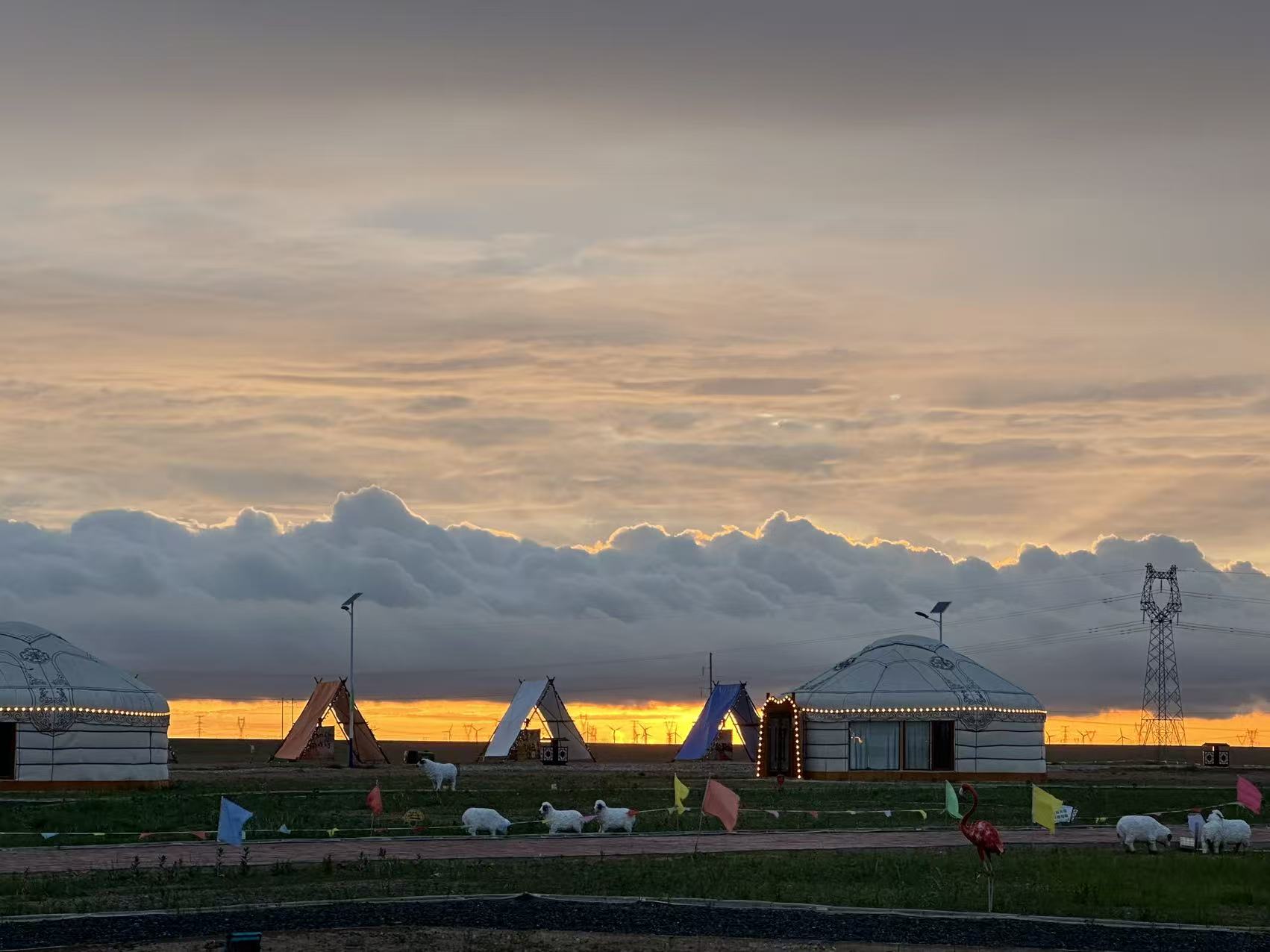}
  \caption{Bayan Gol Starlight Camp in Bayan Gol Gacha, Zhurihe Town, Sunite Right Banner, Inner Mongolia, China. Yurts, tents, and roads with solar streetlights make up the main activity area.}
  \label{fig:site}
\end{figure}

Bayan Gol Starlight Camp is in Bayan Gol Gacha, Zhurihe Town, Sunite Right Banner, Inner Mongolia, China, at 42.448$^\circ$~N, 112.924$^\circ$~E. Its location is shown in \figshortref{fig:site}. The camp hosts stargazing activities and uses solar streetlights along roads and in public areas. Before the retrofit, the light-emitting surfaces of some lamps were directly visible from within the camp. We fitted shields to these lamps and measured the light environment at locations inside and outside the site. The study asks which viewing directions became darker and whether a change could still be measured about 500~m from the camp.

\section{Materials and Methods}

\subsection{Study Site and Lighting Retrofit}

Solar-powered streetlights illuminate roads and public areas at the camp. The lamps are about 4--5~m high, charge from solar panels during the day, and switch on automatically at night. In the pre-retrofit state, the LED light-emitting surfaces were directly visible from several locations inside the camp.

We designed a curved directional shield to fit the lamp housing and its existing mounting points. The directional shield was printed in black PETG. It covered the top and sides of the lamp, remained open at the bottom, and did not cover the solar panel. PETG was selected for an outdoor part that would be screwed to the lamp housing. Previous studies compared the mechanical and thermal properties of printed PETG and PLA and examined water absorption and degradation in both materials~\cite{hsueh2021,moreno2021}. These studies informed the material choice; this project did not test the long-term durability of the directional shield. The black material blocks light through the directional shield itself. The directional shield was attached at the original mounting location.

\begin{figure}[!htbp]
  \centering
  \begin{subfigure}[t]{0.395\linewidth}
    \centering
    \includegraphics[width=\linewidth,height=0.38\textheight,keepaspectratio]{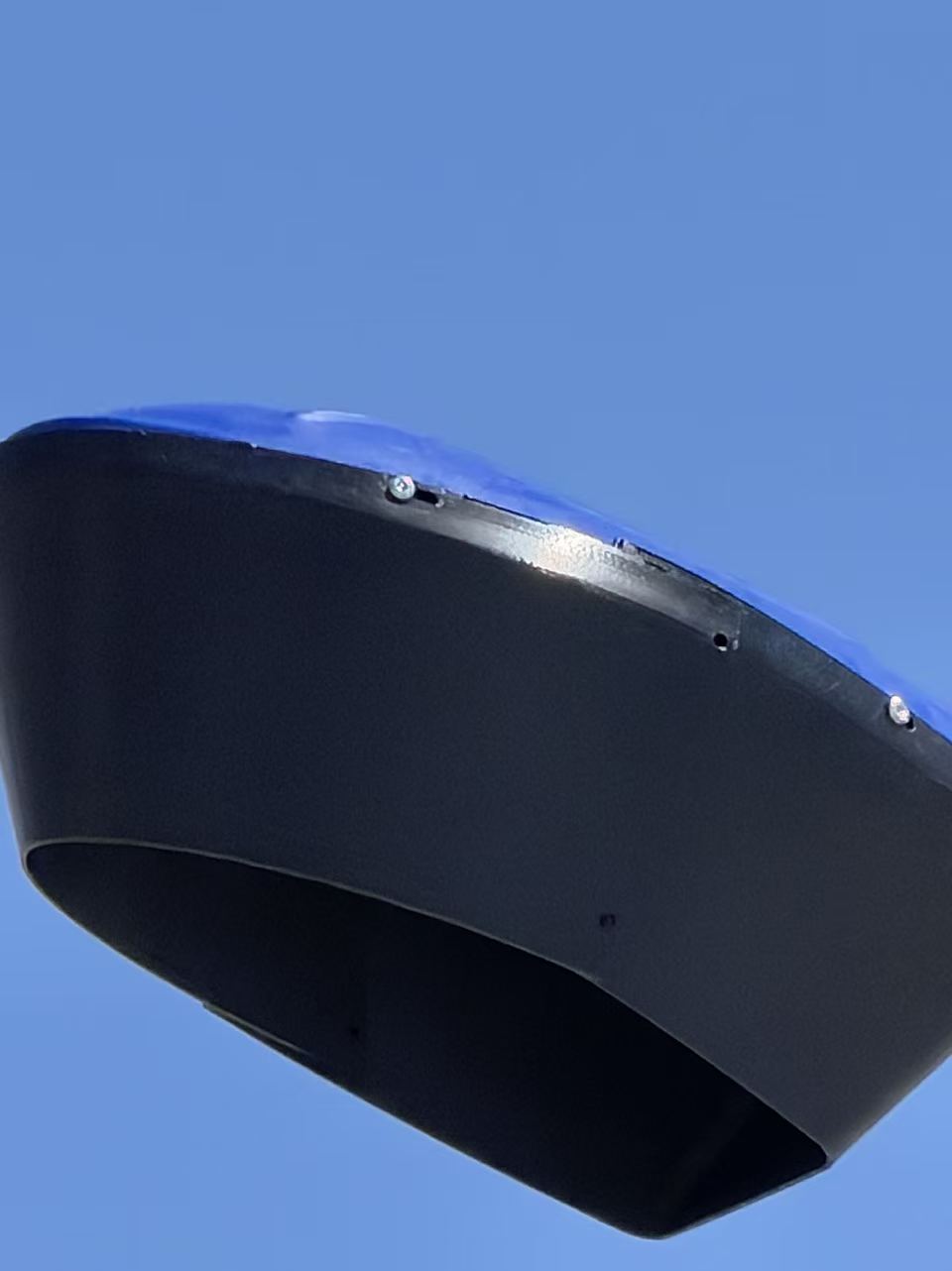}
  \end{subfigure}\hspace{0.035\linewidth}
  \begin{subfigure}[t]{0.395\linewidth}
    \centering
    \includegraphics[width=\linewidth,height=0.38\textheight,keepaspectratio]{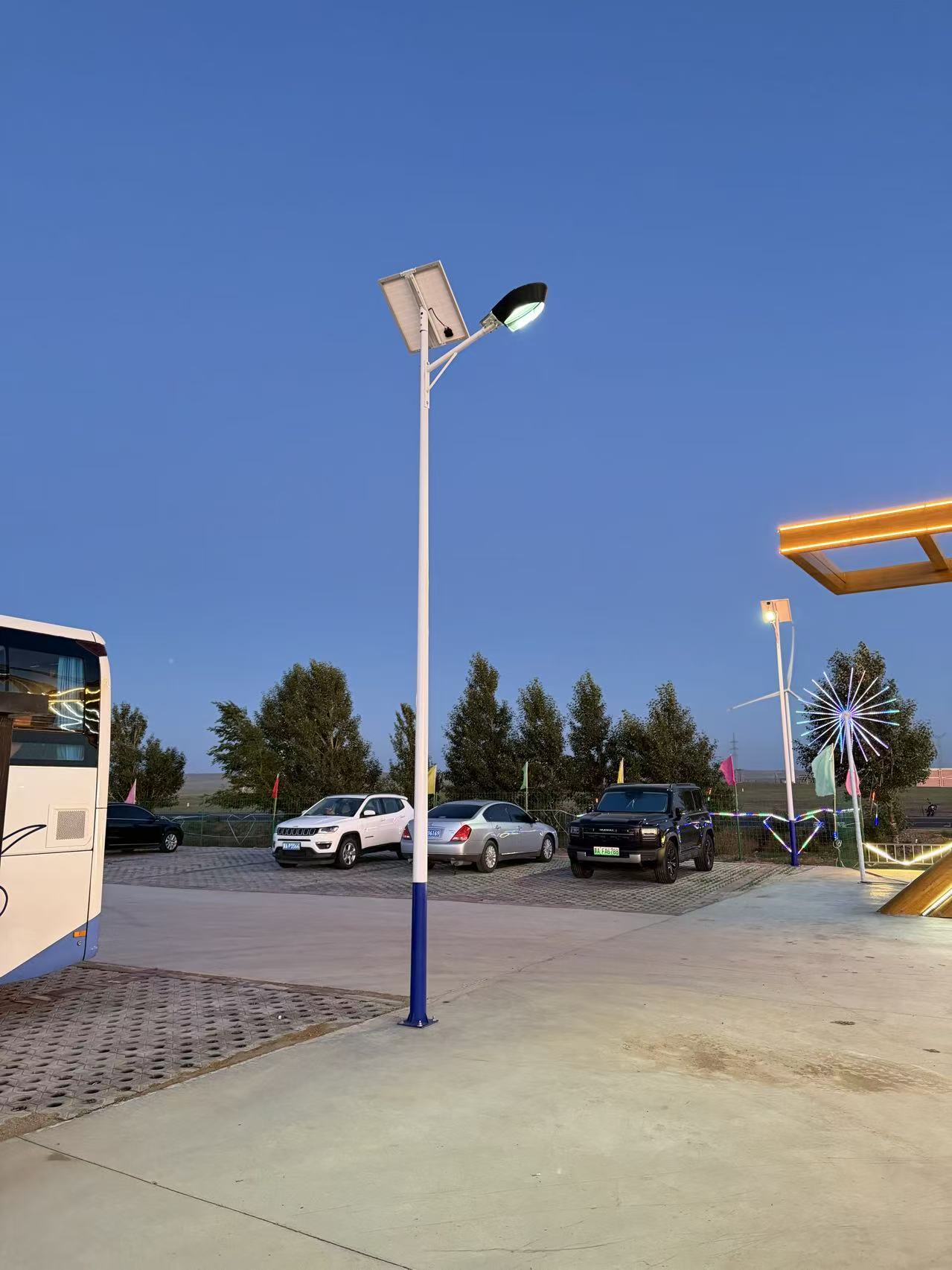}
  \end{subfigure}
  \caption{Three-dimensional printed shield and installation on a solar streetlight. The left panel shows the printed shield, including its side cover and open underside; the right panel shows the shield installed on the streetlight.}
  \label{fig:shield}
\end{figure}

\subsection{SQM Measurements and Photographs}

We took pre-retrofit SQM-L measurements on 14 July 2026 and post-retrofit measurements on 15 July 2026, after installing the directional shields. We used a Unihedron SQM-L sky brightness meter. The SQM-L has an angular response with a full width at half maximum of about 20 degrees and reports readings in magnitudes per square arcsecond~\cite{cinzano2005}. A larger reading indicates a lower measured signal and a darker viewing field.

We fixed the SQM-L to a tripod and aimed it at the planned direction and elevation angle. We pressed the meter button to start each reading and recorded the value when it appeared on the display. An elevation angle of 0 degrees means that the optical axis was horizontal. At this elevation angle, the field of view may include the lower sky, ground, lamps, or buildings, so the reading represents the combined light in that direction. An elevation angle of 90 degrees points toward the zenith~\cite{katz2016,cinzano2005}.

Measurements were made at P1, in the camp center, and P2, about 500~m outside the camp. At P1, we aimed the SQM-L north, east, south, and west. At P2, we aimed it toward the camp, to the observer's left, away from the camp, and to the observer's right; left and right refer to an observer facing the camp. At both sites, measurements were taken at elevation angles of 0, 30, 60, and 90 degrees. Five readings were normally taken for each retrofit state, site, direction, and elevation angle; the value at the 90-degree elevation angle was one zenith reading for each site and retrofit state. For selected conditions with greater short-term variation, seven readings were taken. Before each set, we checked for temporary light sources such as vehicle headlights, flashlights, and nearby activity. We recorded the time, site, retrofit state, direction, elevation angle, and any unusual conditions.

Photographs were taken with a Sony $\alpha$7 III camera and a Sony FE 24~mm F1.4 GM lens. The camera settings were the same in the pre-retrofit and post-retrofit states: f/1.4, ISO 4000, an 8~s exposure, and a color temperature of 3200~K. Images were taken near the camp entrance and parking area. They document the visible lamps, glare, and stars; they are not absolute brightness measurements.

\section{Data Processing}

Five readings were normally used. For selected conditions with seven readings, we removed the highest and lowest readings before calculating the mean and sample standard deviation from the remaining five. The sample standard deviation describes short-term repeatability within one measurement condition; it does not represent uncertainty between the two nights. At P1, we calculated means for the north, east, south, and west directions, then averaged the four directional means at elevation angles of 0, 30, and 60 degrees. At the 90-degree elevation angle, the reading was the single zenith value for that site. At P2, means were calculated separately for the four directions. \tabref{tab:main-results} lists the means used in the main text. \figref{fig:p2} shows the post-retrofit minus pre-retrofit change in each P2 direction.

For a given direction $d$ and elevation angle $e$, the difference between the post-retrofit and pre-retrofit mean readings is given by \eqnref{eq:delta}
\begin{equation}
  \Delta m(d,e) = \overline{m}_{\mathrm{after}}(d,e) - \overline{m}_{\mathrm{before}}(d,e).
  \label{eq:delta}
\end{equation}
A positive $\Delta m$ means that the post-retrofit SQM-L reading is larger and the measured field is darker. Magnitudes use a logarithmic scale, so a difference of 2~mag does not mean a 2\% decrease in brightness. From the magnitude definition~\cite{pogson1856}, the post-retrofit to pre-retrofit equivalent-signal ratio is given by \eqnref{eq:signal-ratio}.
\begin{equation}
  \frac{L_{\mathrm{after}}}{L_{\mathrm{before}}} = 10^{-0.4\Delta m}.
  \label{eq:signal-ratio}
\end{equation}
The percentage decrease relative to the pre-retrofit equivalent signal is given by \eqnref{eq:signal-reduction}
\begin{equation}
  \eta = \left(1 - 10^{-0.4\Delta m}\right)\times 100\%.
  \label{eq:signal-reduction}
\end{equation}
Here, $L$ is the equivalent signal within the instrument's response band for the specified direction. This conversion compares readings only when the same instrument and method are used. The converted percentage applies only to the stated SQM-L field of view and does not represent total lamp output or all-sky brightness.

\begin{table}[!htbp]
  \centering
  \caption{Mean SQM readings before and after the retrofit at the main measurement directions.}
  \label{tab:main-results}
  \small
  \begin{tabular}{@{}llcrrr@{}}
    \toprule
    Site & Direction & Elevation & Before & After & $\Delta$SQM \\
    \midrule
    P1 & Four-direction mean & 0$^\circ$ & 18.184 & 19.807 & $+1.624$ \\
    P1 & Four-direction mean & 30$^\circ$ & 20.210 & 20.785 & $+0.575$ \\
    P1 & Four-direction mean & 60$^\circ$ & 21.064 & 21.039 & $-0.025$ \\
    P1 & Shared zenith reading & 90$^\circ$ & 21.006 & 21.044 & $+0.038$ \\
    P2 & Toward camp & 0$^\circ$ & 18.262 & 20.574 & $+2.312$ \\
    P2 & Toward camp & 30$^\circ$ & 20.702 & 20.888 & $+0.186$ \\
    P2 & Toward camp & 60$^\circ$ & 21.016 & 21.112 & $+0.096$ \\
    P2 & Shared zenith reading & 90$^\circ$ & 21.190 & 21.106 & $-0.084$ \\
    \bottomrule
  \end{tabular}

  \vspace{0.35em}
  \begin{minipage}{\linewidth}
    \footnotesize\textit{Note.} Readings are in \sqmUnit{}. $\Delta$SQM is the post-retrofit mean minus the pre-retrofit mean. Differences were calculated from unrounded means and rounded to three decimal places. Positive values indicate a darker measured field. Each 90$^\circ$ value is one zenith reading for its site.
  \end{minipage}
\end{table}

\section{Results}

\subsection{Photographic Record}

\begin{figure}[!htbp]
  \centering
  \begin{subfigure}[t]{0.31\linewidth}
    \centering
    \includegraphics[width=\linewidth,height=0.30\textheight,keepaspectratio]{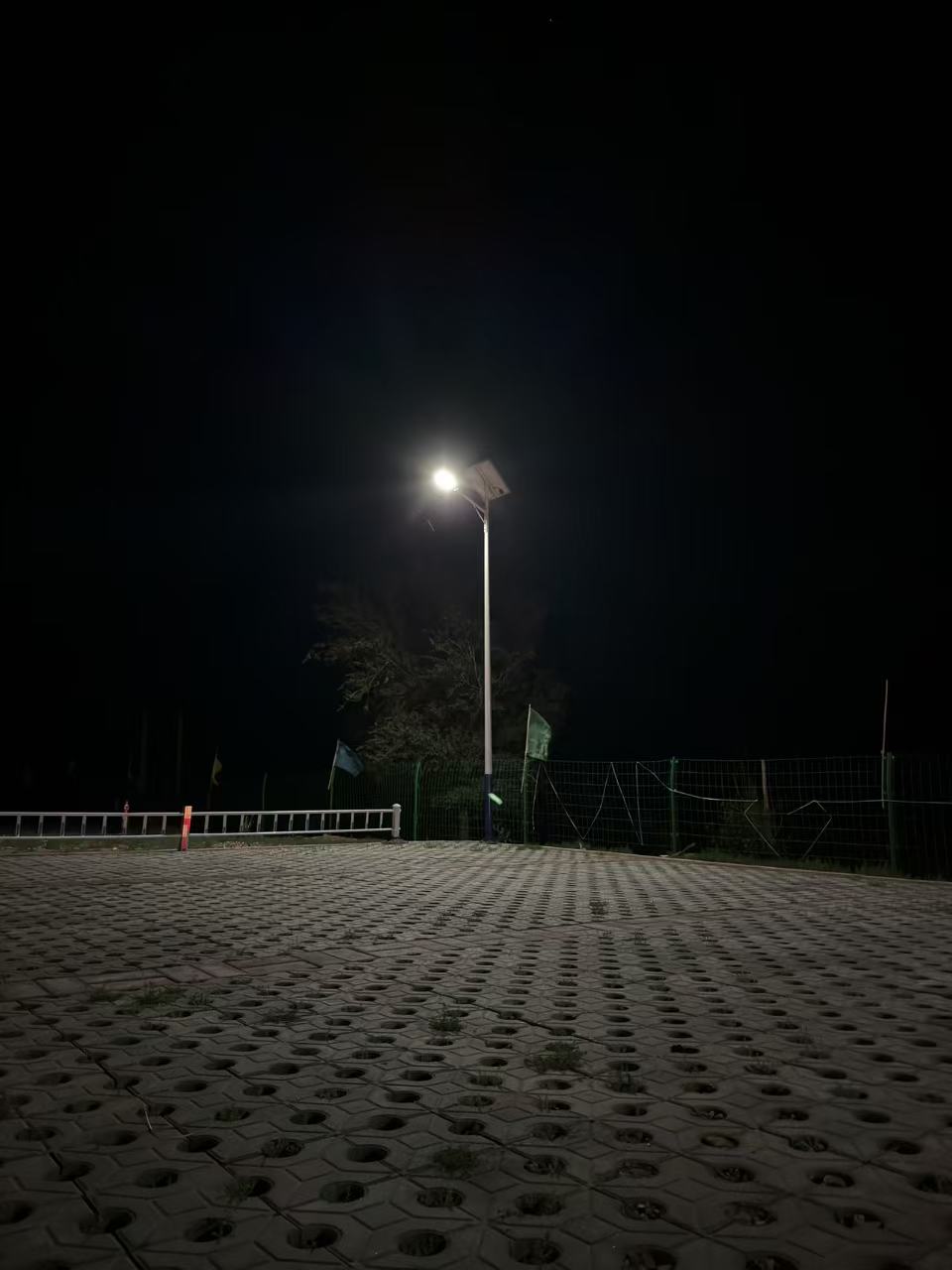}
  \end{subfigure}\hspace{0.025\linewidth}
  \begin{subfigure}[t]{0.31\linewidth}
    \centering
    \includegraphics[width=\linewidth,height=0.30\textheight,keepaspectratio]{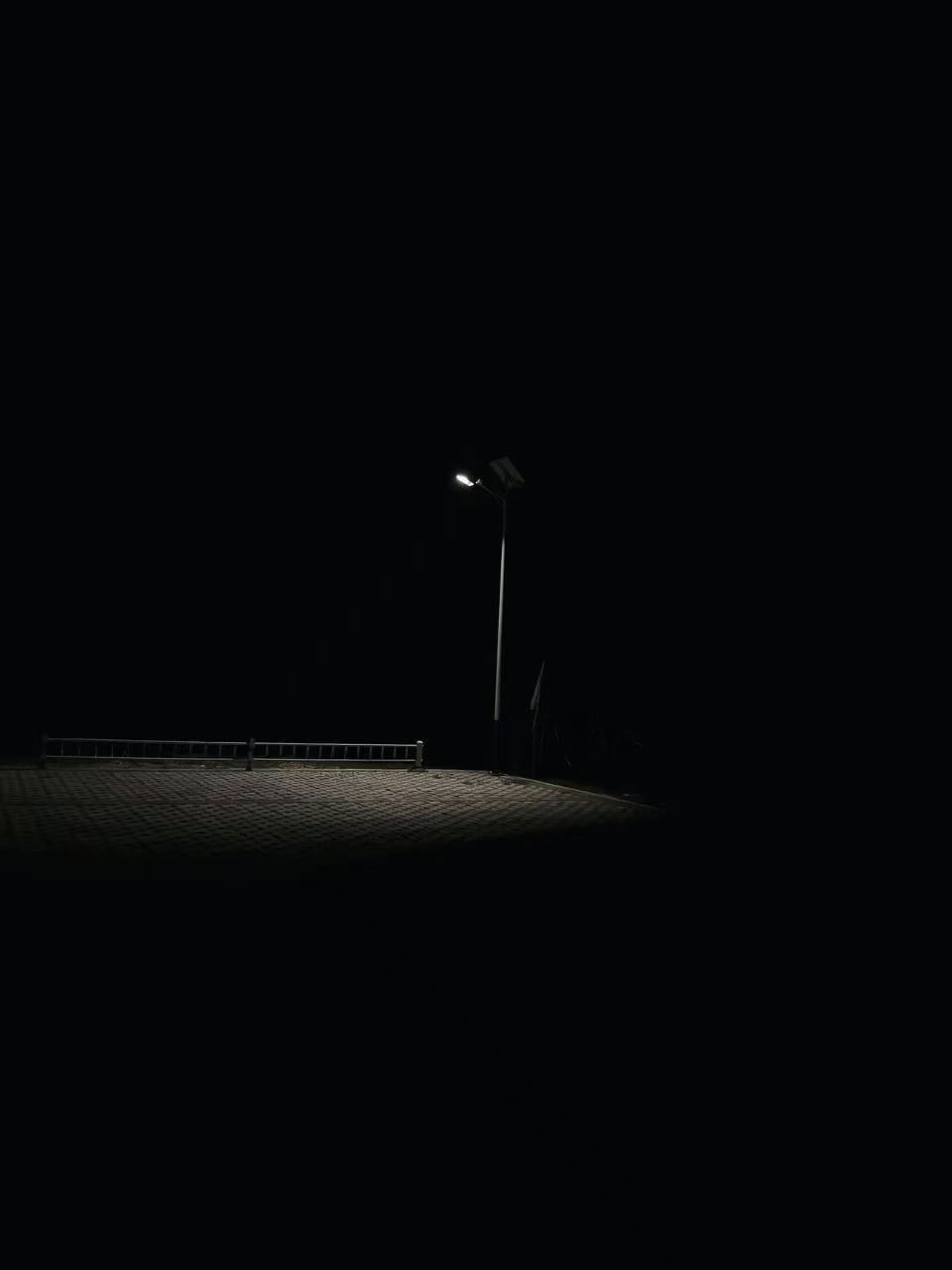}
  \end{subfigure}
  \caption{Nighttime appearance of the same type of streetlight before (left) and after (right) installation of the shield.}
  \label{fig:lamp-photos}
\end{figure}

\begin{figure}[!t]
  \centering
  \makebox[\linewidth][c]{%
    \begin{subfigure}[t]{0.44\linewidth}
      \centering
      \includegraphics[width=\linewidth,height=0.21\textheight,keepaspectratio]{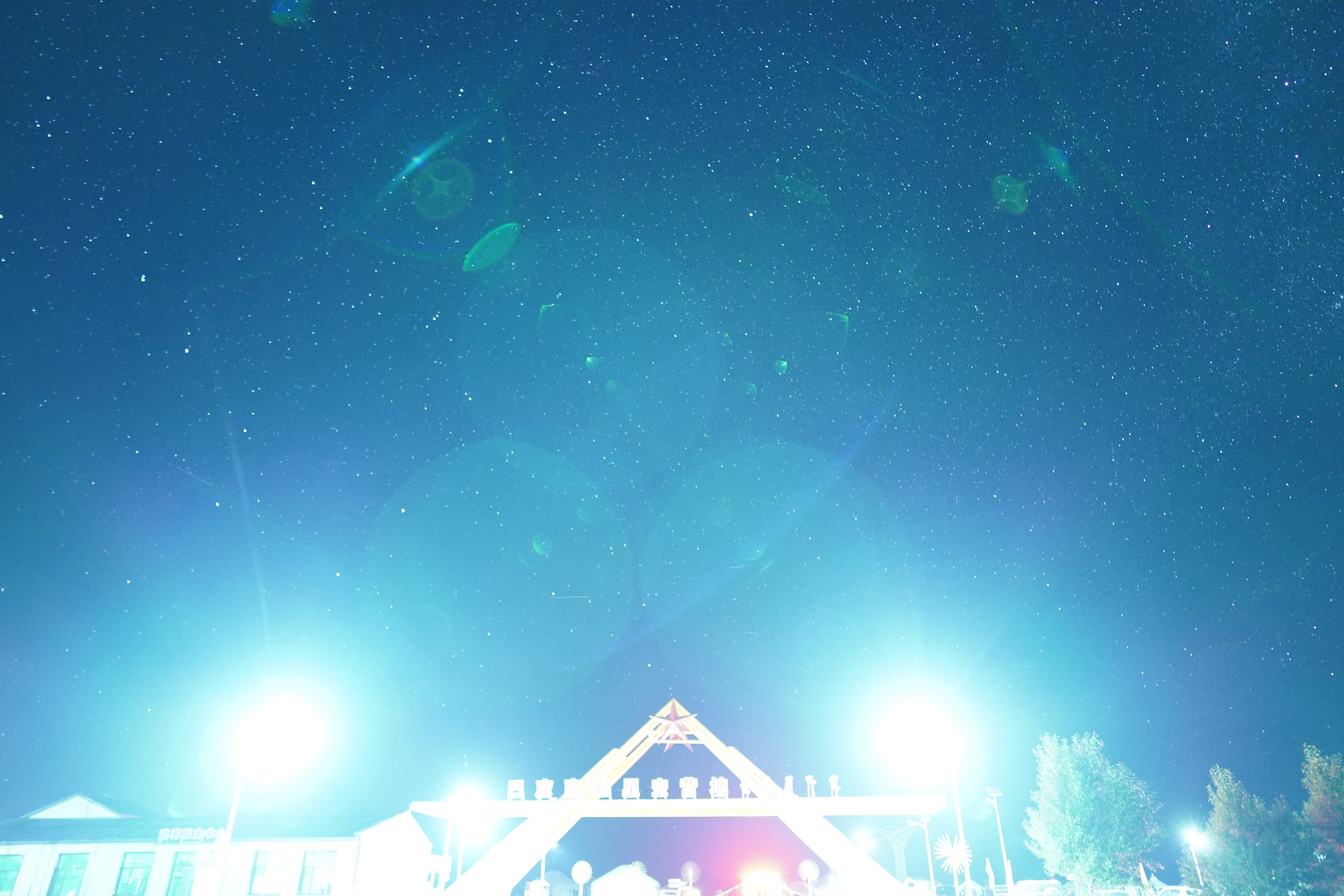}
    \end{subfigure}\hspace{0.015\linewidth}%
    \begin{subfigure}[t]{0.44\linewidth}
      \centering
      \includegraphics[width=\linewidth,height=0.21\textheight,keepaspectratio]{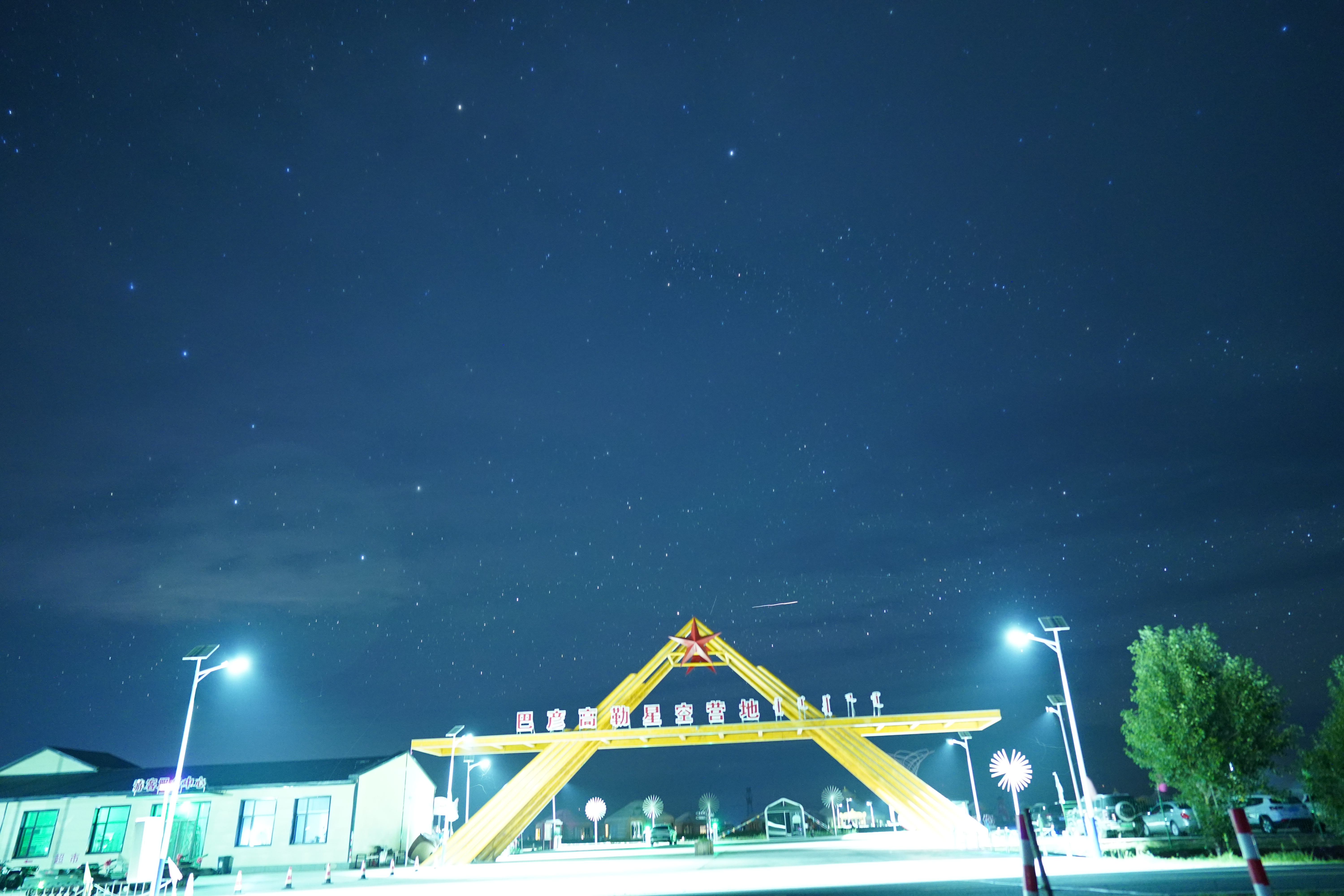}
    \end{subfigure}%
  }

  \vspace{0.8em}
  \makebox[\linewidth][c]{%
    \begin{subfigure}[t]{0.44\linewidth}
      \centering
      \includegraphics[width=\linewidth,height=0.21\textheight,keepaspectratio]{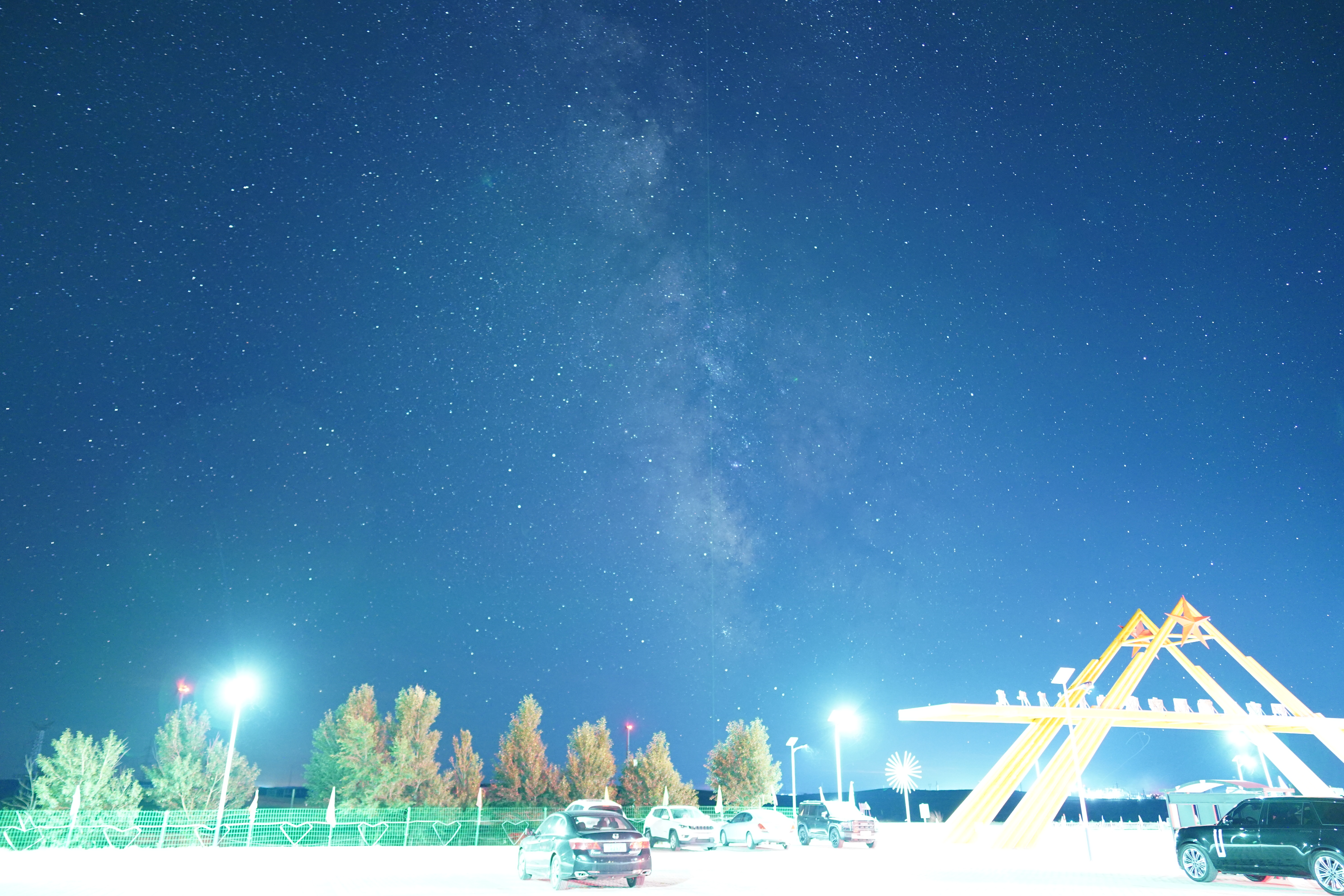}
    \end{subfigure}\hspace{0.015\linewidth}%
    \begin{subfigure}[t]{0.44\linewidth}
      \centering
      \includegraphics[width=\linewidth,height=0.21\textheight,keepaspectratio]{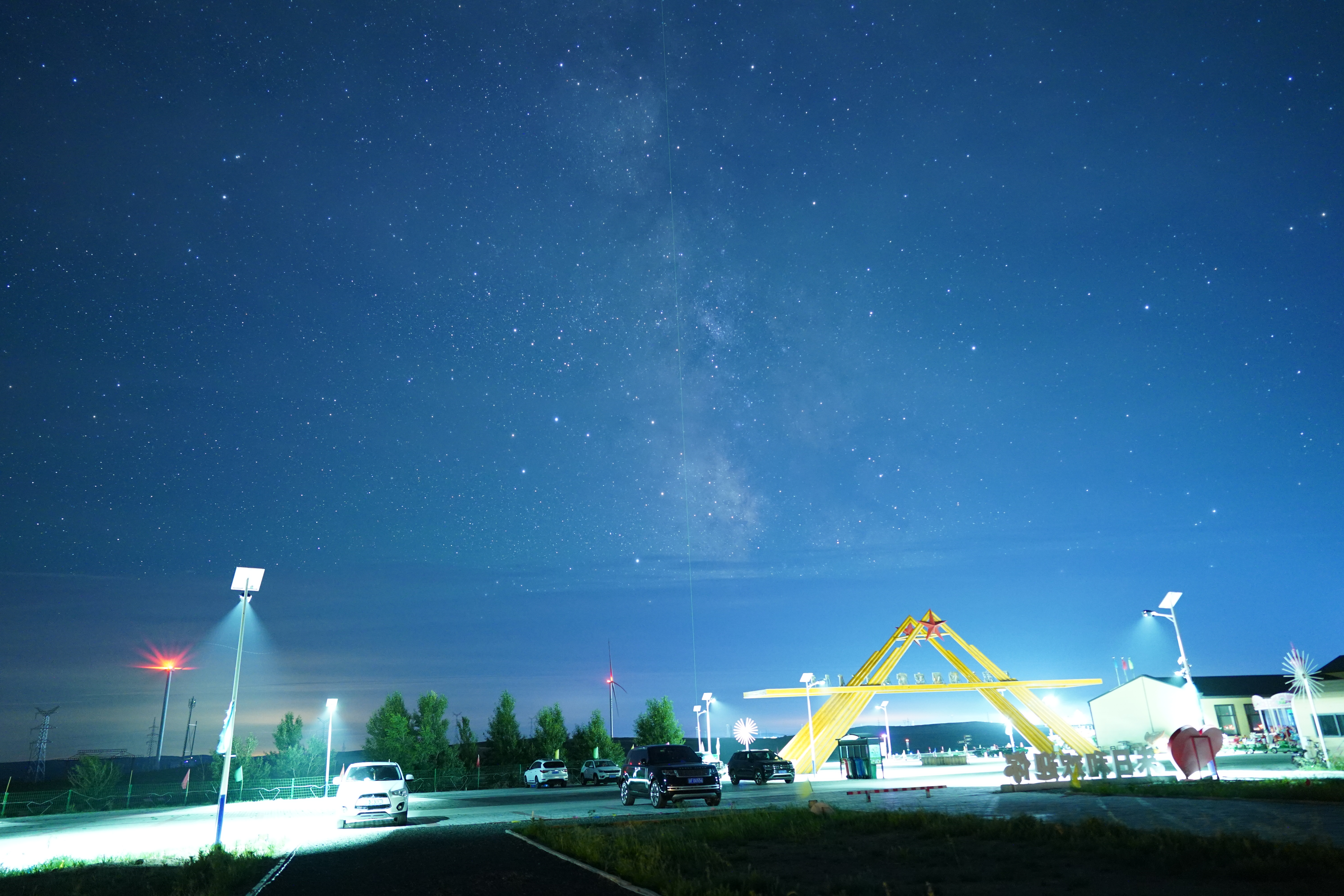}
    \end{subfigure}%
  }
  \caption{Nighttime lighting near the camp entrance in the upper row and parking area in the lower row. Images on the left were taken before installation; images on the right were taken after installation. The same camera and exposure settings were used in both conditions.}
  \label{fig:camp-photos}
\end{figure}

\figref{fig:lamp-photos} records the visible light distribution around the same type of streetlight before and after the directional shield was installed. In the pre-retrofit image, the LED surface is directly visible from the side, and a broad bright region surrounds the lamp. In the post-retrofit image, the shield blocks the side view of the light-emitting surface. The illuminated area is more concentrated below the fixture, while the pavement beneath the lamp remains visible.

\figref{fig:camp-photos} presents two additional comparisons made with the same camera and exposure settings. The upper row shows the camp entrance. Before the retrofit, bright regions around the two lamps occupy a large part of the lower image, and several areas of lens flare are visible against the sky. After the retrofit, the bright regions around the lamps appear smaller, and the outlines of the buildings and background stars are more readily distinguishable.

The lower row of \figref{fig:camp-photos} shows the parking area. In both images, the road, parked vehicles, and nearby structures remain visible. The post-retrofit image contains smaller bright regions around the lamps and less extensive foreground glare. The two locations therefore show the same qualitative pattern: the visible light around the fixtures is more spatially confined after installation, while the illuminated ground remains visible. These photographs provide a visual record only. They were not photometrically calibrated and were not used to calculate changes in brightness.

\subsection{Measurements at P1}

\begin{figure}[!htbp]
  \centering
  \begin{subfigure}[t]{0.48\linewidth}
    \centering
    \includegraphics[width=\linewidth]{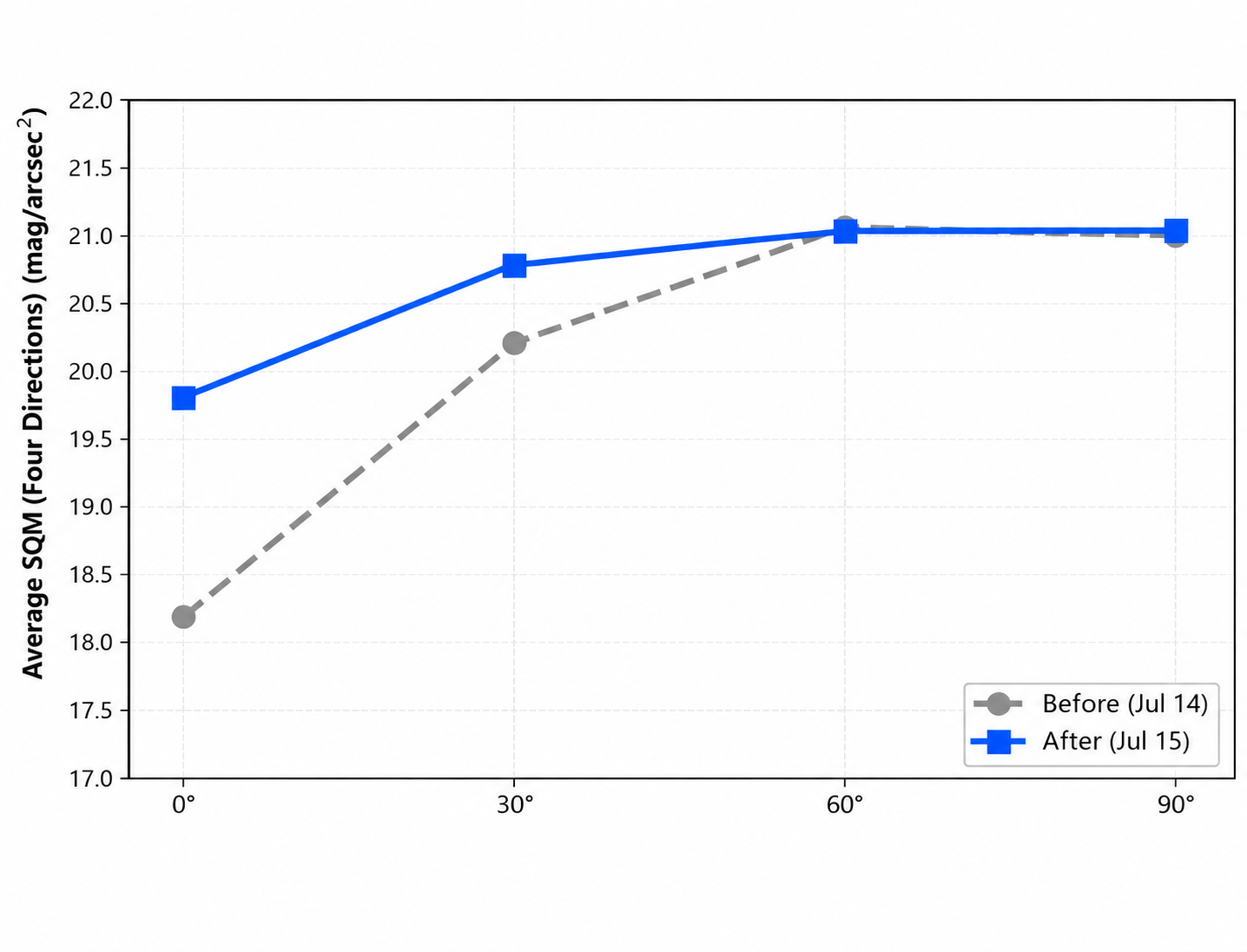}
  \end{subfigure}\hfill
  \begin{subfigure}[t]{0.48\linewidth}
    \centering
    \includegraphics[width=\linewidth]{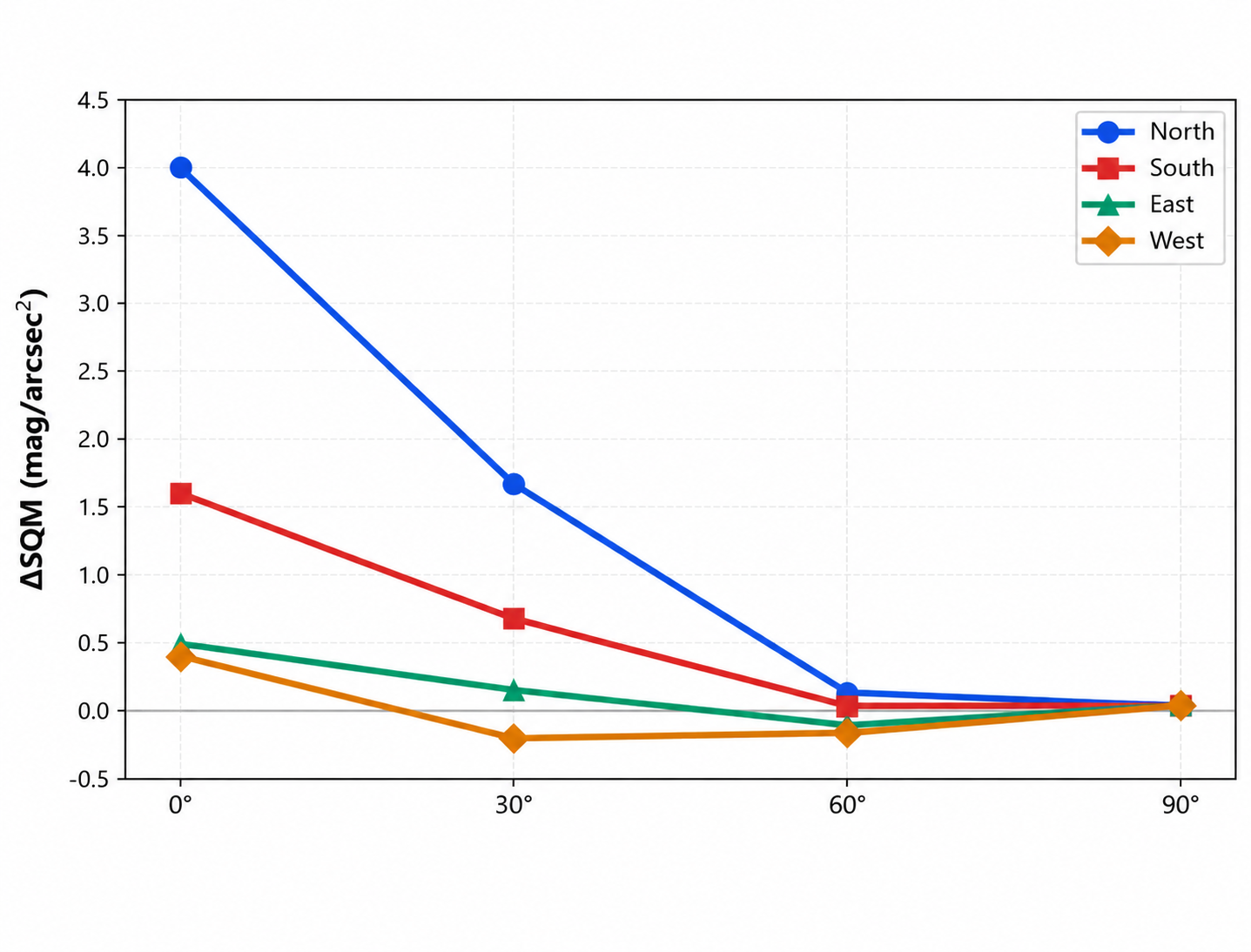}
  \end{subfigure}
  \caption{SQM changes at P1. The left panel shows mean SQM readings by elevation; the 0$^\circ$, 30$^\circ$, and 60$^\circ$ values are four-direction means, and 90$^\circ$ is the shared zenith reading. The right panel shows direction-specific $\Delta$SQM; the black point marks the shared zenith reading. Positive values indicate a darker measured field after installation.}
  \label{fig:p1}
\end{figure}

P1 was located near the center of the camp and was measured in the north, east, south, and west directions. The left panel of \figref{fig:p1} compares the four-direction mean at each elevation, while the right panel shows the direction-specific changes. The largest mean difference occurred in the horizontal field. At 0$^\circ$, the mean SQM reading increased from 18.184 to 19.807~\sqmUnit{}, giving a $\Delta$SQM of $+1.624$~\sqmUnit{}. According to \eqnref{eq:signal-reduction}, this change corresponds to an approximately 77.6\% decrease in the equivalent SQM signal within the measured field.

Elevation-dependent differences were smaller. At 30$^\circ$, the four-direction mean rose from 20.210 to 20.785~\sqmUnit{}, with a $\Delta$SQM of $+0.575$~\sqmUnit{}. At 60$^\circ$, the mean changed from 21.064 to 21.039~\sqmUnit{}, giving a $\Delta$SQM of $-0.025$~\sqmUnit{}. The zenith reading rose from 21.006 to 21.044~\sqmUnit{}, with a $\Delta$SQM of $+0.038$~\sqmUnit{}. In \figref{fig:p1}(a), the pre- and post-retrofit curves are farthest apart at 0$^\circ$, less separated at 30$^\circ$, and nearly overlap at 60$^\circ$ and 90$^\circ$.

The horizontal change also depended on direction. At 0$^\circ$, $\Delta$SQM was positive in all four directions, but its magnitude was uneven. The changes were $+4.000$~\sqmUnit{} to the north, $+1.602$~\sqmUnit{} to the south, $+0.492$~\sqmUnit{} to the east, and $+0.400$~\sqmUnit{} to the west. The north-facing reading increased from 15.016 to 19.016~\sqmUnit{} and produced the largest directional difference at P1. At higher elevations, the changes did not have a consistent sign. The west-facing values were $-0.202$~\sqmUnit{} at 30$^\circ$ and $-0.164$~\sqmUnit{} at 60$^\circ$, while the east-facing value at 60$^\circ$ was $-0.106$~\sqmUnit{}.

\begin{figure}[!htbp]
  \centering
  \begin{subfigure}[t]{0.42\linewidth}
    \centering
    \includegraphics[width=\linewidth]{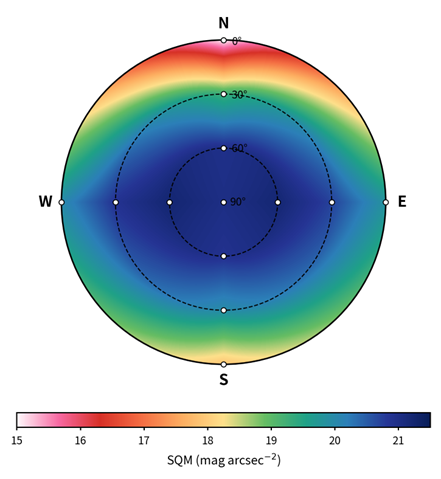}
  \end{subfigure}\hfill
  \begin{subfigure}[t]{0.42\linewidth}
    \centering
    \includegraphics[width=\linewidth]{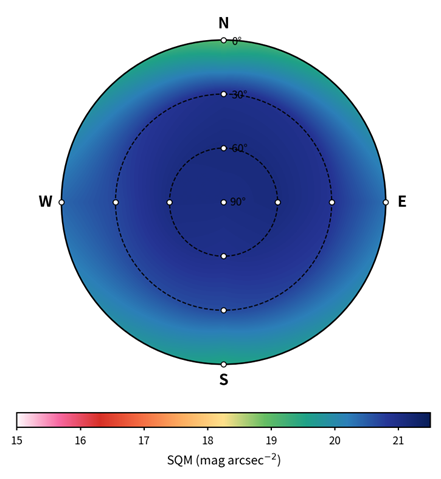}
  \end{subfigure}
  \caption{Circular interpolation of mean SQM readings at P1 before installation (left) and after installation (right). The same color scale is used in both panels. White points mark sampled directions; the outer edge represents a horizontal 0$^\circ$ view and the center represents the 90$^\circ$ zenith.}
  \label{fig:p1-maps}
\end{figure}

\FloatBarrier
\figref{fig:p1-maps} displays the P1 readings as two circular interpolations with the same color scale. The white points mark the measured directions and elevations; colors between these points are interpolated and do not represent additional measurements. In the pre-retrofit panel, the lowest SQM readings occupy a pronounced region near the northern edge, which represents the north-facing horizontal field. This low-value region is smaller in the post-retrofit panel. By comparison, the colors near the center of the two panels, corresponding to the zenith, remain similar. The spatial display therefore reproduces the directional and elevation-dependent pattern reported in \figref{fig:p1}.

\subsection{Measurements at P2}

\begin{figure}[!htbp]
  \centering
  \includegraphics[width=0.85\textwidth]{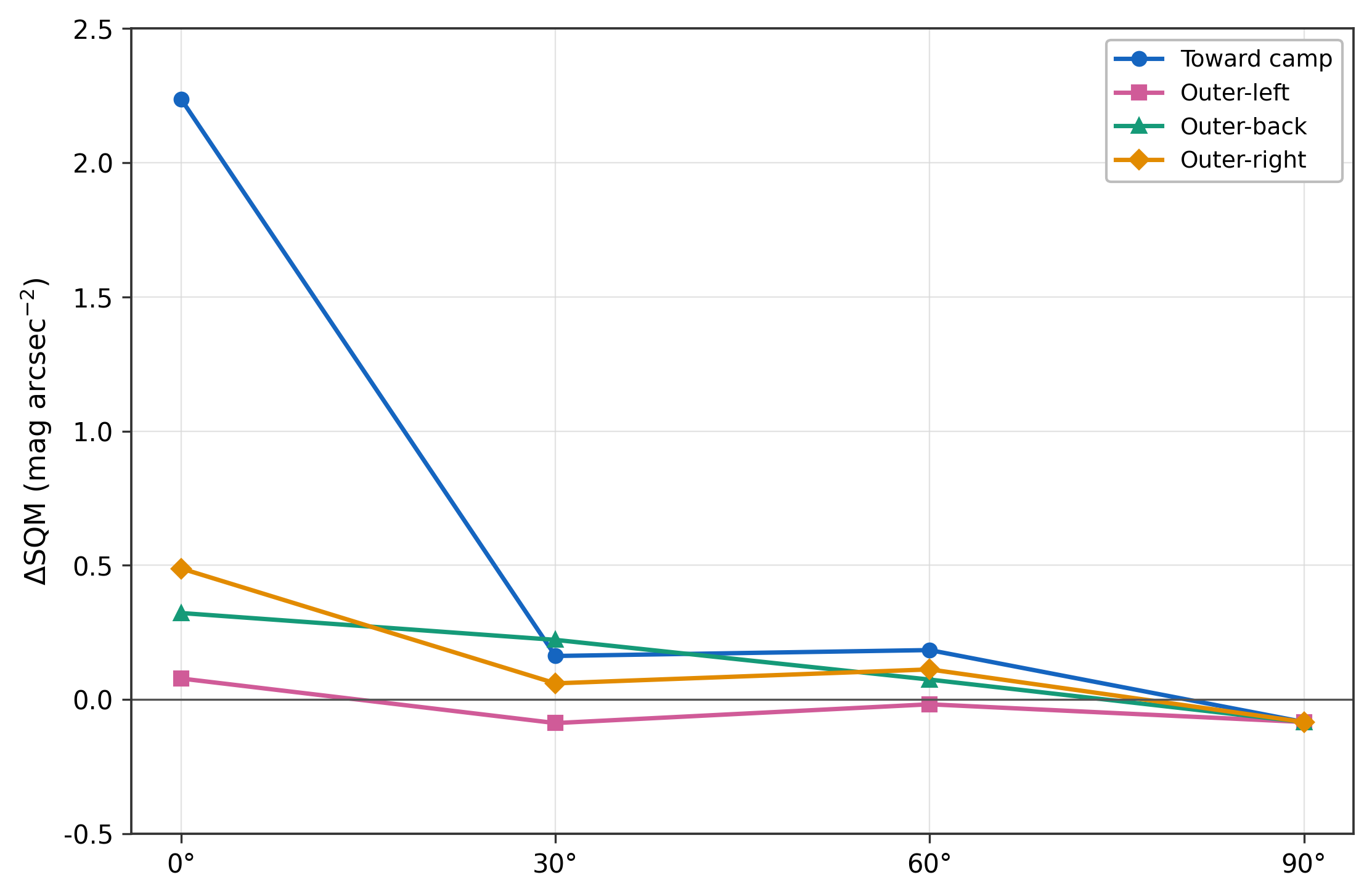}
  \caption{Change in SQM readings in four directions at P2. ``Toward camp,'' ``Outer-left,'' ``Outer-back,'' and ``Outer-right'' indicate the four viewing directions. The 90$^\circ$ values are the same shared zenith reading; the lines are connected to that point for display only. Each change is the post-retrofit mean minus the pre-retrofit mean.}
  \label{fig:p2}
\end{figure}

P2 was located approximately 500~m outside the camp. \figref{fig:p2} compares the post-retrofit minus pre-retrofit changes in the direction toward the camp and in the three surrounding directions. The largest difference occurred in the horizontal field toward the camp. At 0$^\circ$, the mean reading in this direction increased from 18.262 to 20.574~\sqmUnit{}, giving a $\Delta$SQM of $+2.312$~\sqmUnit{}. \eqnref{eq:signal-reduction} converts this value to an approximately 88.1\% decrease in the equivalent SQM signal within that viewing field.

The other horizontal directions changed by smaller amounts. At 0$^\circ$, $\Delta$SQM was $+0.154$~\sqmUnit{} to the observer's left, $+0.398$~\sqmUnit{} away from the camp, and $+0.564$~\sqmUnit{} to the observer's right. All three values were below the $+2.312$~\sqmUnit{} measured toward the camp.

The differences toward the camp were also smaller at higher elevations. At 30$^\circ$, the mean reading increased from 20.702 to 20.888~\sqmUnit{}, with a $\Delta$SQM of $+0.186$~\sqmUnit{}. At 60$^\circ$, it increased from 21.016 to 21.112~\sqmUnit{}, with a $\Delta$SQM of $+0.096$~\sqmUnit{}. The shared zenith reading changed from 21.190 to 21.106~\sqmUnit{}, giving a $\Delta$SQM of $-0.084$~\sqmUnit{}. The four lines in \figref{fig:p2} meet at this common 90$^\circ$ value because the zenith was represented by one shared reading rather than four direction-specific measurements. Across the P2 measurements, the largest post-retrofit difference was therefore confined to the horizontal view toward the camp.

\FloatBarrier

\section{Discussion}

\subsection{Contribution and practical significance}

Stargazing camps need lighting for paths and public areas, but directly visible lamps can produce glare and interfere with low-elevation observing. This study tested a retrofit that retained the existing solar-powered fixtures and changed only the direction in which their light was visible. The work connects three stages that are often treated separately: shield design, installation under field conditions, and direction-resolved evaluation after installation. Matched-exposure photographs documented the change in visible glare, while SQM-L measurements quantified how the observed light field changed with azimuth and elevation.

The direction-resolved measurements are a central feature of the study. The clearest response occurred in horizontal and low-elevation fields, whereas the zenith changed little at both sites (Figures~\ref{fig:p1}--\ref{fig:p2}). A zenith-only survey would therefore have recorded little of the improvement most relevant to an observer looking across the camp. The measurement geometry was matched to the expected function of the shield, which was to limit lateral and oblique light rather than to alter every part of the sky. This result supports the use of defined fields of view and pointing directions when evaluating local lighting interventions~\cite{hanel2018}.

The two measurement sites extend the assessment beyond the immediate vicinity of the lamps. P1 described the light environment within the camp. P2, located about 500~m away, tested whether a camp-facing change could still be detected from a potential observing area outside the main lighting zone. The response at P2 shows that the effect of a local retrofit can be relevant beyond the installation site itself. It also shows why evaluations of stargazing facilities should include the places where visitors may actually observe, rather than relying only on a central or zenith measurement.

The retrofit offers a practical workflow for existing lighting systems. Lamps with visible emitting surfaces can first be identified from observing and walking areas. Shields can then be fitted to these lamps. Repeating measurements along the same lines of sight provides a direct check of whether the intervention changed the intended viewing fields. This approach is consistent with the principle that outdoor light should be directed toward the area where it is needed~\cite{darksky2020}. It may be useful for camps, small observatories, nature-education sites, and other locations that need both nighttime access and dark-sky conditions.

Another practical advantage is the low implementation cost. The shields were mounted on the existing lamps without replacing the fixtures or modifying the electrical system, which reduced both material use and installation work.

\subsection{Interpretation of the directional changes}

The shield geometry helps explain the concentration of the response at low elevations. Its upper and side panels reduce the area of the emitting surface visible from horizontal and oblique directions, while the open underside preserves downward illumination. Less direct lamp light and stray light should therefore enter these viewing fields after installation. Artificial-light propagation near outdoor sources also depends on the emission pattern, source-observer geometry, and atmospheric scattering~\cite{aube2015,bara2023}.

Differences among azimuths are expected in a real camp environment. The number, orientation, distance, and visibility of lamps were not the same in all directions around P1. Buildings and other foreground objects may also have blocked or reflected light. Comparable direction-dependent variation in SQM measurements has been reported~\cite{katz2016}. The present azimuthal pattern therefore reflects the arrangement of lamps and sightlines at this site, rather than a uniform response across the camp.

The 0$^\circ$ readings describe a combined visual field. Depending on direction, the SQM-L field may contain lower sky, lamp surfaces, illuminated ground, buildings, and reflected light. These readings are useful for assessing the light environment experienced along a specific line of sight, but they are not measurements of sky background alone. The equivalent signal reductions reported in the Results apply to the measured fields and should not be extended to the full sky hemisphere.

The photographs and SQM-L readings provide complementary evidence. The photographs show how the visible extent of lamp glare changed and whether the illuminated ground remained visible. The SQM-L measurements provide a repeatable numerical comparison for selected viewing fields. Neither method identifies all components of the light field by itself, but their agreement supports the interpretation that the shields primarily changed lateral and low-elevation exposure.

The retained downward illumination is important for the intended use of the camp. Roads and parking areas remained visible in the post-retrofit photographs, suggesting that shielding did not simply remove all useful light. Photographs, however, cannot determine whether the illumination was adequate or uniform. Horizontal illuminance on the road, vertical illuminance at pedestrian eye level, and discomfort glare should be measured before the retrofit is evaluated against formal lighting requirements.

\subsection{Improvements and further work}

The main experimental limitation is that the before-and-after measurements were obtained on different nights. Cloud cover, aerosols, atmospheric transparency, and temporary light sources can alter nighttime brightness between sessions~\cite{kyba2015,jechow2017}. Five or seven repeated readings describe short-term variation under one condition, but not variation between nights. Future measurements should cover several comparable moonless nights and record cloud cover, visibility, humidity, wind, measurement time, and temporary light sources. If the installation can be changed safely, alternating shielded and unshielded states during the same night would provide a more direct comparison.

The spatial sampling should also be expanded. One outside site is sufficient to show that a camp-facing change was detectable at that location, but it cannot define how the response decreases with distance. Several outside sites should be placed along comparable sightlines at increasing distances, with foreground obstructions documented at each position. A calibrated all-sky camera used alongside the SQM-L could determine whether the change extends beyond the selected measurement fields~\cite{jechow2018}. Quantitative image analysis would require fixed acquisition settings, correction for lens response and vignetting, and photometric calibration.

Further work should evaluate lighting performance and material durability together. Road illuminance, pedestrian-level vertical illuminance, and glare should be measured before and after shielding. Long-term monitoring should also record deformation, discoloration, cracking, loose fasteners, and dust accumulation on the PETG shields under ultraviolet exposure, temperature cycling, wind, and precipitation. These measurements would show whether the improvement in the observing environment can be maintained without reducing the required lighting function.

\section{Conclusion}

We fitted PETG directional shields to existing solar streetlights at Bayan Gol Starlight Camp in Bayan Gol Gacha, Zhurihe Town, Sunite Right Banner, Inner Mongolia, China, and compared readings from 14 and 15 July 2026. At P1 in the camp center, the mean horizontal change across four directions was $+1.624$~\sqmUnit{}, equivalent to an approximately 77.6\% decrease in the SQM signal within that field. At P2, about 500~m outside the camp, the horizontal reading toward the camp increased from 18.262 to 20.574~\sqmUnit{}, a change of $+2.312$~\sqmUnit{} and an approximately 88.1\% decrease in the equivalent SQM signal within that field. The largest changes occurred in low-elevation views and varied by direction.

These measurements indicate that directional shielding can change light entering selected viewing fields while retaining existing lamps. They apply to the sampled fields and do not establish reductions in total lamp output, all-sky brightness, or light pollution at other locations. Because the comparison covered two nights, repeated measurements under comparable conditions are needed. Directional shields can be evaluated as retrofits at stargazing sites using fixed viewing directions. Because the retrofit retains the existing lamps and electrical systems, it also provides a relatively low-cost and practical approach to dark-sky protection at stargazing sites.

\printbibliography[title={References}]

\end{document}